\documentclass[
reprint,
longbibliography,
amsmath,amssymb,
aps,
prl,
floatfix,
]{revtex4-2}

\usepackage{graphicx}%
\usepackage{xcolor}
\usepackage{cancel}
\usepackage{dcolumn}%
\usepackage{bm}%
\usepackage{booktabs}
\usepackage{comment}

\begin{document}

\title{Probing bilayer topological order with layer-resolved transport}
\author{Hongquan Liu$^{1,2}$, J.I.A. Li$^{3}$, and D. E. Feldman$^{1,2}$}
\affiliation{$^1$Department of Physics, Brown University, Providence, Rhode Island 02912, USA}
\affiliation{$^2$Brown Theoretical Physics Center, Brown University, Providence, Rhode Island 02912, USA}
\affiliation{$^{3}$Department of Physics, University of Texas at Austin, Austin, TX 78712, USA}

\date{\today}

\begin{abstract}
Shot noise has been used to measure fractional charges of anyons. The value of the charge imposes constraints on fractional statistics but does not determine it. 
This {\color{black}limitation} is particularly important in multi-component systems.
For example, the zero charge of neutral anyons in bilayer graphene  gives no information about their statistics at all. We propose a {\color{black}way to shed light on} the statistics of charged and neutral anyons in multi-component systems with layer-resolved or spin-resolved noise. The protocol applies to the fractional quantum spin Hall effect in MoTe$_2$, topological states in multi-layer graphene and bilayer GaAs, and to recently discovered fractional excitons in bilayer graphene. The approach relies on the relation between statistics and the distribution of the anyon charge over the components. Information about statistics can also be extracted from a simpler measurement of the layer-resolved electric current through a narrow constriction in a Hall bar
even in the presence of long-range interactions and other non-universal effects.
\end{abstract}

\maketitle


{\color{black} Fractional charge and fractional statistics are two key features of topological matter \cite{review-FH}. Anyonic interferometry \cite{manfra20,nakamura2023:fabry,Willett2023interference, kundu2023:mach-zehnder,chiral2024-1,chiral2024-2,werkmeister2025interference,Samuelson2024interference,ronen-5/2-interference,kim-5/2-interference} has emerged as a tool to probe statistics while noise has long been used to measure fractional charges  \cite{de1997direct,saminadayar1997observation,reznikov1999observation,dolev2008observation}.} At millikelvin temperatures, Nyquist noise provides a powerful and elegant approach to thermometry \cite{Jezoin,texp1} 
and helps detect tiny thermal conductance of the edges of topological liquids \cite{Jezoin,texp1,texp2,texp3}.
It was proposed that a combination of noise and interferometry could serve as a probe of fractional statistics  \cite{feldman2007:shot_noise}. Striking noise signatures of statistics were identified in Mach-Zehnder interferometry \cite{zucker2016}. Anyonic colliders gave evidence of anyonic statistics via noise \cite{rosenow2016current,col2,col3}. Their physics can be understood in terms of quasiparticle interference in the time domain \cite{time-domain}.


{\color{black}Electric noise alone, on the other hand, has not been shown to reveal much about statistics, though it has interesting applications in probing topological modes in the spin Hall effect \cite{Colin2025,Colin2023,Colin2017} and Laughlin states \cite{Martin2020,Gu2022}. The observation of a fractional charge necessarily implies anyonic statistics, but charge alone does not determine  what the statistics are  \cite{MZ-review}. {\color{black} An extreme example of this is neutral quasiparticles, which have been demonstrated recently in quantum Hall graphene bilayers ~\cite{3/8+3/8,Zhang2025fractionalexciton}. The fact that their charge is zero} sheds no light on statistics at all.}

{\color{black} The challenge is particularly acute in multi-component systems, where anyonic interferometry has not been implemented.
{\color{black}Yet multi-component systems have recently attracted considerable interest in the field of topological matter.} 
Several topological liquids have been found in bilayer graphene and in bilayer quantum wells in GaAs \cite{bilayer-1/2_1,bilayer-1/2_2,Suen1994,Shabani2013,Eisenstein2014,Zibrov2017,Li.17b,Ghahari}.  More recently, the putative fractional quantum spin Hall effect was observed in {\color{black}tMoTe$_2$} bilayers \cite{FSHE}, and neutral anyonic excitons were discovered in bilayer graphene devices \cite{Zhang2025fractionalexciton}. In all such systems, excitation charge is distributed over layers or spin projections, and the resulting fractional statistics depends on this distribution. {\color{black} Conveniently,} probing multi-component electronic order with layer-resolved technique is well established across different material platforms, such as semiconductor quantum wells and graphene heterostructures ~\cite{Kellogg2002exciton,Kellogg2004exciton,Tutuc2004exciton,Nandi2012exciton,Eisenstein2014,Li2017exciton,Liu2017exciton,Liu2022crossover,Zhang2025fractionalexciton,3/8+3/8}.} {\color{black} Yet these techniques have not been used to extract anyon statistics.} 

{\color{black}} In this paper, we find that overcoming the limitations of noise probes in bilayer and other multi-component systems helps  distinguish the statistics of identically charged anyons {\color{black} by revealing the structure of anyons that constrains statistics.} Moreover, information about the statistics can be extracted from a simpler measurement of the layer-resolved electric current, for properly chosen values of the voltage bias in the layers.
{\color{black}We will show that the charge distribution  can be determined through the component-resolved shot noise, if different  driving voltages are applied to the layers. We will also show that a simpler measurement of the component-resolved current sheds light on the statistics, even if long-range interactions, dissipation, and other effects make the I-V curve non-universal. In some cases, the layer-resolved currents and voltages have opposite signs.} 

Below we use a well-known example of the quantum Hall effect in half-filled bilayer systems \cite{bilayer-1/2_1,bilayer-1/2_2} to illustrate the idea.
Related physics is likely present in the recently discovered $3/8+3/8$ state in bilayer graphene \cite{3/8+3/8}. Numerical work \cite{331-1,331-2} suggests  an Abelian 331 state \cite{Halperin1983} in the limit of weak inter-layer tunneling or no tunneling. Another possible candidate is a non-Abelian Pfaffian state \cite{Moore1991}. 
{\color{black} To help distinguish between these candidates,} we {\color{black} find} that  I-V curves and shot noise in a constriction in a  Hall bar exhibit qualitative differences in Abelian and non-Abelian states. The differences are explained, in part, by different charge distributions between the layers for Abelian and non-Abelian anyons. Another effect is less intuitive: interlayer correlations affect the spatial dependence of the electric field in a way sensitive to fractional statistics. {\color{black} More broadly}, we observe that the same idea allows probing the nature of the putative fractional spin Hall state in {\color{black}tMoTe$_2$}. Finally, we show how a modification of the idea can be used to probe neutral anyons in graphene bilayers \cite{Zhang2025fractionalexciton}. Many other systems can be probed in a similar way. {\color{black} In all cases, we assume that a component such as a layer or valley is a good quantum number everywhere in the sample. In GaAs and graphene bilayers this means negligible inter-layer tunneling. In tMoTe$_2$ this assumes that inter-valley scattering is unimportant.}

{\color{black} We consider anyon tunneling through a constriction in Fig. 1a in the 331 state. To understand transport, we need to understand the topological order and edge theory.}
The 331 topological order is described by the $K$-matrix \cite{WenBook}

\begin{equation}
K=\begin{pmatrix}
3 & 1\\
1 & 3
\end{pmatrix}
\end{equation}
and the charge vector $t=(1,1)$, where the two entries of the vector correspond to the two layers. This encodes both the statistics and the edge theory with the Lagrangian

\begin{equation}
L=\int dx \sum_{i,j=1}^2\frac{1}{4\pi}(\partial_t\phi_iK_{ij}\partial_x\phi_j-
\partial_x\phi_iU_{ij}\partial_x\phi_j),
\end{equation}
where the charge densities in the two layers are $e\partial_x\phi_i/2\pi$, and $U_{ij}$ depends on the mode velocities  and interactions. 
We can ignore the tunneling between the layers since the tunneling term in the Lagrangian, $\int dx [\xi(x)\exp(2i[\phi_1(x)-\phi_2(x)]) +{\rm h.c.}]$, is irrelevant in the renormalization group sense. Since all edge modes are co-propagating, the scaling dimensions of quasiparticle operators do not depend on $U_{ij}$. Quasiparticles are described by integer vectors $q=(n,m)$ and carry layer resolved charges
of $e(1,0)K^{-1}{\color{black}{q^T}}$ and $e(0,1)K^{-1}{\color{black}{q^T}}$.

We consider a single constriction between two edges (Fig. 1a), which allows quasiparticle tunneling. 
One can apply two different voltage biases $V_{1,2}$ to the two layers.
This will drive the tunneling of not only charged but also neutral quasiparticles, which carry opposite charges in the layers.
The tunneling is described by the contribution to the Hamiltonian of the form $\Gamma\hat T+\Gamma^*\hat T^\dagger$, where the tunneling operator ${\color{black}\hat T=\exp[i(\phi_1^u+\phi_1^d,\phi_2^u+\phi_2^d)q^T]}$ transfers a quasiparticle between the two edges,
and the superscripts $u$ and $d$ label the edges above and below the constriction. The most relevant tunneling operators in the limit of low voltages and temperatures transfer particles of types $p_1=(1,0)$ and $p_2=(0,1)$ with the layer-resolved charges of $(3e/8,-e/8)$ and
$(-e/8,3e/8)$ respectively. The scaling dimension of the tunneling operators $g_c=p_iK^{-1}p_i^T=3/8$.
While the most relevant operators determine the dominant tunneling process in the limit of vanishing temperature and voltage across the constriction, the tunneling of lowest-charge quasiparticles is a competing process at finite temperatures \cite{review-FH} since lower-charge quasiparticles are expected to have a lower energy gap and hence a lower tunneling barrier through the bulk. The lowest-charge particles are electrically neutral. The most relevant such particles are $(1,-1)$ and $(-1, 1)$ with the corresponding scaling dimension of $g_n=1$ and the layer resolved charges $\pm e/2$. $g_n$ is considerably higher than $g_c=3/8$. Nevertheless, we will consider both neutral and charged anyons. 

{\color{black} To compute the observable signatures, we now define}
the layer current operator $\hat I_k={\color{black}\dot {\hat{Q}}_k}=i[\hat H, \hat Q_k]$, where $\hat H$ is the Hamiltonian and $\hat Q_k$ is the charge operator of the upper edge in layer $k$. We find $\hat I_k=ie_k\Gamma \hat T-ie_k\Gamma^*\hat T^\dagger$, where $e_k$ is the layer-resolved charge of the tunneling quasiparticle. We will focus on the limit of low temperatures and voltages such that $eV>T$. In the absence of tunneling, the four charges of the two layer-resolved edge channels on the two sides of the constriction {\color{black} are separately conserved}. Thus, in the limit of weak tunneling, we can assign four chemical potentials to the four channels. We will set them at zero {\color{black}below} the constriction and at $eV_{1,2}$ {\color{black}above} the constriction. A convenient way to incorporate them into the tunneling problem involves \cite{law2006:PhysRevB.74.045319} the use of {\color{black} the} interaction representation such that the Hamiltonian is shifted by $\sum Q_s\mu_s/e$, where $Q_s$ and $\mu_s$ are the charges and the chemical potentials of the four channels. This introduces time-dependence {\color{black} to} the tunneling operators and hence to the current operator, {\color{black}$\hat T\rightarrow \hat T\exp(-ie\tilde V t)$},
 where $e\tilde V=e_1V_1+e_2V_2$.

 The average current 

 \begin{equation}
 I_k=\langle S(-\infty, 0)\hat I_k(0)S(0,-\infty)\rangle,    
 \end{equation}
where ${\color{black}S={\rm T}\exp(-i\int [\Gamma \hat T(t)+\mathrm{h.c.}] dt)}$ is the $S$-matrix in the interaction representation and the angular brackets represent the thermal average. In the lowest order of perturbation theory in powers of $\Gamma$, the expression reduces to

\begin{equation}
I_k=e_k|\Gamma|^2\int_{-\infty}^0 dt\langle [\hat T(t),\hat T^\dagger(0)]-[\hat T^\dagger(t),T(0)] \rangle,
\end{equation}
where {\color{black}$T(t)\sim\exp(-ie\tilde Vt)$}. 
The expansion is legitimate as long as $\Gamma T^{g-1}$ is small, where $T$ is the temperature and $g$ is the scaling dimension of the tunneling operator. 
If several equally relevant quasiparticle tunneling operators are present, contributions, quadratic in each of the tunneling operators, should be added. 

We first consider charged anyons with $g=g_c=3/8$. In our case, $(1,0)$ and $(0,1)$ are equally relevant quasiparticles. To single out the contribution from only one quasiparticle type, we consider
$V_1=3V_2.$ The effective voltage, experienced by the particles of type $(0,1)$ becomes $e\tilde V_{(0,1)}=-3V_2\cdot e/8+V_2\cdot 3e/8=0$. Hence, in the leading order, the current of the $(0,1)$ quasiparticles is the same as at zero voltage, that is, zero. The effective voltage for $(1,0)$ quasiparticles becomes $e\tilde V_{(1,0)}=eV_2$, and their current is nonzero. This leads us to the first prediction: the ratio of the tunneling currents in the two layers reflects the fractions of a quasiparticle charge in each layer:

\begin{equation}
\label{-3}
\frac{I_1}{I_2}=\frac{3e/8}{-e/8}=-3.
\end{equation}
The ratio is negative even though the voltages in each layer have the same sign.
{\color{black} This counter-intuitive behavior can be understood by observing that charges $-e/8$ experience 
two oppositely directed forces from the in-layer electric field and from the $3e/8$-charged component of the quasiparticle. Weak fields cannot break quasiparticles, so both components are forced to move in the direction of the sum of the two in-layer electric forces.}
{\color{black} This robustness contrasts with that of a more commonly used diagnostic:} the exponent in the power dependence of the tunneling conductance on the temperature was proposed as a probe of statistics. Unfortunately, the observed exponents are non-universal due to long-range Coulomb interactions and other mechanisms \cite{review-FH}. We note that the prediction (\ref{-3}) is insensitive to those effects.

More refined information can be extracted from shot noise \cite{MZ-review}, which tells what the layer distribution of the charge of each quasiparticle is. The noise is defined as

\begin{equation}
S_k=\int_{-\infty}^{\infty} dt [\langle\hat I_k(t)\hat I_k(0)+\hat I_k(0)\hat I_k(t)\rangle-2\langle\hat I_k(0)\rangle^2].
\end{equation}
At $eV_{1,2}> T$, the noise follows the Schottky formula
$S_k=2|e_kI_k|$. From this formula the layer-resolved charges of $3e/8$ and $-e/8$ can be read out. Recently, a technique for probing cross-correlations of currents in fractional quantum Hall systems was developed \cite{cross-correlation} and provides a higher accuracy in fractional charge experiments than the auto-correlation noise \cite{cross-correlation-theory}. The cross-correlation noise 


\begin{equation}
    \begin{aligned}
        S_{12}= \quad &  \\
\frac{1}{2}\int_{-\infty}^{\infty}\!\!\! dt&[\langle \hat I_1(t)\hat I_2(0)+\hat I_2(0)\hat I_1(t)+\hat I_1(0)\hat I_2(t)+\hat I_2(t)\hat I_1(0) \rangle   \\ 
&-4\langle \hat I_1\rangle\langle \hat I_2 \rangle]={\color{black}-|2e_2I_1|=-|2e_1I_2|}
    \end{aligned}
\end{equation}
can be used as evidence of simultaneous tunneling of layer-resolved charges $e_{1,2}$.

As discussed above, it may happen that the dominant tunneling process involves $(1,-1)$ particles and their $(-1,1)$ antiparticles. The analysis is similar in that case. The current and noise only depend
on the difference of the biases applied to the layers: $\tilde V=(V_1-V_2)/2$. The layer-resolved currents are opposite, $I_1=-I_2$, and the Fano factors in the noise all equal $e/2$: $S_1=S_2=-S_{12}=|eI_1|$.

We now turn to another possibility, a non-Abelian Pfaffian state 
\cite{Moore1991} in a bilayer. {\color{black} See a detailed discussion of that and all the other states we mention in Supplemental Material \cite{SM}.} {\color{black} This state is well-known in the single-layer limit \cite{Moore1991}. Its minimal anyon carries charge $e/4$. The edge theory is described by the Lagrangian
\begin{equation}
\label{new-1}
 \mathcal{L} = 
 i\psi(\partial_t+v_n\partial_x)\psi-\frac{2}{4\pi}\partial_x\phi(\partial_t+v_c\partial_x)\phi,
\end{equation}
where $\psi$ is a neutral Majorana fermion with the speed $v_n$, the charge density is $\frac{e\partial_x\phi}{2\pi}$, and $v_c$ is the speed of the charged mode.} 

{\color{black} Although numerics in the limit of weak inter-layer tunneling favor the 331 state \cite{331-2}, we analyze the Pfaffian state as an alternative candidate to see what distinct transport signatures it would produce.}
In the Pfaffian state, half of the electrons are in each layer, but the wave function is essentially single-layer, if one can neglect the distance between the layers. This is equivalent to a single-layer state in which electrons are randomly divided into two equal groups called layers. At first sight, this might create rich noise physics, since the quasiparticle charge of $e/4$ can be divided between the layers in an arbitrary proportion. However, an attempt to repeat the above calculations fails, since the edge contains an electrically neutral chiral Majorana channel and a single charged mode, {\color{black} equation (\ref{new-1})}. There is no way to include two chemical potentials in that model. This happens because the Pfaffian state can be thought of as a superconductor of composite fermions \cite{t2,tlho1995}. Each composite fermion carries charge $e$ and has two intra-layer and two inter-layer fluxes of the magnetic field attached, so that a Cooper pair creates four fluxes in each layer. In a half-filled system, the fluxes bind the total charge of $-4\times{\frac{1}{2}}\times e=-2e$, which is equally distributed across the layers. Hence, if the two composite fermions in the Cooper pair are taken from the same layer, the electrically neutral pair has non-zero charges $e$ and $-e$ in the two layers. It follows that the in-plane electric-field component  in the top layer must be exactly the same as the in-plane component of the field in the bottom layer at the point immediately below (Fig. 2). Otherwise, Cooper pairs experience net electric force and charges rearrange. 

This issue is absent in the 331 state even though it can also be understood as made of Cooper pairs. The reason is that all Cooper pairs are made of particles from the two different layers in the 331 state \cite{tlho1995}. Hence, the pair has zero charge in each layer.

{\color{black} We now address transport signatures of the Pfaffian state.} We see that the state can be understood as an exciton superfluid. Such systems exhibit the counterflow effect \cite{Eisenstein2014} (Fig. 2). If the two layers are electrically connected on one side of a Hall bar, then the current entering on the other side in one layer returns through the other. The charge moves through the system without resistance as bound pairs of opposite charges in the two layers. 
If a voltage bias is applied to a constriction, we expect {\color{black} the absolute values of the} drain currents and {\color{black}noises} to be divided equally between the two layers {\color{black}(See Supplemental Material for details)}. {\color{black}The relations between the currents and the noises are imposed by symmetry and may look paradoxical if bias is applied to one layer only in the absence of interlayer tunneling. The paradox is explained by neutral superfluid currents of Cooper-pair dipoles. The currents reshuffle the charge between the modes.}

We now turn to the putative fractional spin Hall state \cite{fsqh-1,sodemann,fsqh-2,fsqh-3,fsqh-4} in bilayer twisted MoTe$_2$ at the filling factor $\nu=3$. In this system, spin projection is conserved and plays the same role as the layer index in the previous discussion \cite{mote2-review}. Spin-up and -down electrons have opposite chirality on the edge separating $\nu=2$ from $\nu=3$. Thus, they can be separately biased by two sources on the right and left of the constriction (Fig. 1b) and analyzed separately in two drains connected to the upper edge. {\color{black}We assume weak interaction of contra-propagating modes \cite{SM}.}

{\color{black} We wish to distinguish candidate states with layer-resolved transport. We review the candidates below and, in more detail, in Supplemental Material. In our discussion of the proposed states, we ignore the two filled Chern bands.}
The simplest scenario for the observed state is a time-reversal-invariant topological liquid made of two copies \cite{fsqh-3,fsqh-4} of opposite-chirality $\nu=1/2$ liquids of the 16-fold way \cite{ma2016-16}, {\color{black} such as two Pfaffian liquids with the edge action (\ref{new-1}) for one spin projection and a similar action with the opposite mode propagation direction for the other spin projection. All such states have the same component-resolved charges of their minimal anyons: $(e/4,0)$ and $(0,e/4)$. Another candidate is the minimal topological order \cite{fsqh-2}. Its minimal excitations have charge $e/2$ equally distributed over the components. Its edge structure includes two contra-propagating charged Bose-modes and no neutral modes. It turns out that electrons are gapped on the edge \cite{fsqh-2}.}

Recent evidence of time-reversal-symmetry breaking \cite{time-reversal} is also consistent with another proposal \cite{sodemann} {\color{black} of PH-$mmn$ state}, which generalizes the 331 state and is described by a three-component $K$-matrix

\begin{equation}
K=\begin{pmatrix}
m & n & 0\\
n & m & 0\\
0 & 0 & -1
\end{pmatrix},
\end{equation}
where $n$ is odd, $m=n+2$, and the spin-up and -down charge vectors $q_\uparrow=(1,0,0)$ and
$q_{\downarrow}=(0,-1,1)$. The state can be understood as the $mmn$ state of spin-up electrons and spin-down holes. {\color{black}This proposal is attractive since it explains the absence of a robust Hall plateau in experiment.} {\color{black}Indeed, some of the gapped excitations in that state are itinerant \cite{sodemann}.} {\color{black} Anyons are described by vectors $t=(a,b,0)$ with integer $a$ and $b$. The component charges are $e(ma-nb,na-mb)/(m^2-n^2)$. We will see that spin-resolved transport in the constriction geometry shows dramatically different signatures in this PH-$mmn$ state and the other proposed states.

To understand transport in the geometry of Fig. 1b we need to figure out which particles dominate tunneling across the QPC. At low voltages and temperatures, the tunneling current depends on the bare tunneling amplitude and the scaling dimension of the tunneling operator. Higher tunneling amplitudes are expected for lower-charge excitations, such as neutral anyons. Thus, we need to understand the structure of the neutral anyons as well as anyons with the lowest scaling dimension of the tunneling operator in each of the states. The scaling dimensions depend on the details of the edge theory, which is well understood for the states of the 16-fold way and the minimal order \cite{fsqh-3,fsqh-2}. The edge theory of the PH-$mmn$ state is developed in the End Matter. In particular, equation (\ref{new-2}) can be used to estimate the scaling dimensions of tunneling operators.} 

{\color{black} We next analyze the transport through the constriction in the PH-$mmn$ state. We will then compare with the states of the 16-fold way and the minimal order. First, we} identify the most relevant tunneling operators across the constriction. At $n>1$, {\color{black} according to equation (\ref{new-2}),} such operators {\color{black}\cite{footnote-neutral}} transfer neutral quasiparticles  ${\color{black}(1,1,0)}$ and have the scaling dimension $g=1/(n+1)$ in the absence of  interaction {\color{black} between contra-propagating channels} (see the {\color{black}Supplemental Material} for the interacting case). {\color{black} Since these operators are both the most relevant and transfer the minimal possible total charge of 0, it is likely that they dominate tunneling at $n>1$.} The spin resolved charges of the ${\color{black}(1,1,0)}$ quasiparticles are $\pm e/2(n+1)$. At $n=1$, the most relevant operators transfer charged quasiparticles ${\color{black}(1,0,0)}$ and ${\color{black}(0, 1,0)}$. Their spin-resolved charges are
$(3e/8, e/8)$ and $(e/8, 3e/8)$ respectively.
In the absence of interaction {\color{black}between contra-propagating channels}, the scaling dimension is the same $g=3/8$ for both quasiparticle types. Interactions corrections make them different (see the {\color{black}Supplemental Material}). 

{\color{black} We thus see two possible candidates to dominate the tunneling transport at $n=1$: neutral excitations with the scaling dimension of the tunneling operator $g_n=1/(n+1)=1/2$ and charged excitations with the scaling dimension $g_c=3/8$. The bare amplitude is likely higher for neutral anyons while the two scaling dimensions are close. Hence,}
it is likely that the tunneling of neutral ${\color{black}(1,1,0)}$ quasiparticles remains the dominant process at $n=1$ as long as the voltage and the temperature are not very low. We thus start with the review of the ${\color{black}(1,1,0)}$ tunneling {\color{black} at all $n$}. Similar to the previous discussion, $I_\uparrow({\color{black}V=}V_\uparrow-V_\downarrow)=-I_\downarrow$ and 
$S_{\uparrow\uparrow}=S_{\downarrow\downarrow}=-S_{\uparrow\downarrow}={\color{black}|eI_{\uparrow}/(n+1)|}$. 
{\color{black}The relation of the tunneling currents $I_{\uparrow,\downarrow}$ with the drain currents and noises is addressed in Ref. \onlinecite{SM}.}
At $n\ne 1$, this is clearly different from all states of the 16-fold way {\color{black} and the minimal order}, where the minimal layer-resolved quasiparticle charge \cite{fsqh-3,ma2016-16,fsqh-3} is $e/4$ so that the noise is ${\color{black}|eI_{\uparrow}/2|}$.

{\color{black} $n=1$ requires separate analysis. One can apply the same bias to both spin channels. Shot noise would show the minimal anyon charge $e/4$ in the states of the 16-fold way and the charge of $e/2$ in the PH-331 state. The same minimal anyon charge of $e/2$ is predicted in the minimal topological order from Ref. \onlinecite{fsqh-2}. One can distinguish it from the PH-331 state by looking at the cross-correlation noise between the two drains on the upper edge in Fig. 1 when the same bias is applied to both spin channels. Crucially, anyons (1,0,0) and (0,1,0) make identical contributions to the cross-correlation noise, which is proportional to the product of the charge components $3e/8$ and $e/8$, so that the noise $S_{331}=2\frac{I}{e/2}\times \frac{3e}{8}\times\frac{e}{8}=\frac{3eI}{16}$, where $I$ is the total tunneling current. The most relevant operators of an $e/2$ anyon in the minimal topological order create equal charges $e/4$ in the two channels, so that the cross-correlation noise $S_m=\frac{eI}{4}$. Another way to detect the minimal order is based on tunneling into its edge through a region at the filling factor $\nu=2$. In contrast to the PH-331 state, the minimal topological order does not allow electron tunneling into its edge so that charge will tunnel in the form of charge-$2e$ pairs.}

The same ideas apply in our final example. Recent experiments give evidence of bound pairs of opposite charges, i.e., excitons, in bilayers with two different or  identical Jain states in the layers \cite{Zhang2025fractionalexciton,3/8+3/8}. 
{\color{black} We focus on two layers in the Laughlin $1/3$ state with the filling factors $\nu_1$ and $\nu_2$ such that $\nu_1+\nu_2=2/3$. Neutral excitons made of two anyons of charge $\pm e/3$ in the two layers are introduced in the system when $\nu_1\ne\nu_2$.}
No bulk transport was observed near the center of the plateau {\color{black} at $\nu_1=\nu_2=1/3$}, and only neutral-exciton transport was observed slightly away from the center of the plateau \cite{3/8+3/8}. This suggests that excitons will also dominate tunneling through a constriction near the center of the plateau {\color{black} at a small $\nu_1-\nu_2$}. In this case, the tunneling current and noise depend only on the difference of the layer-resolved voltages, and $I_1=-I_2$, $S_{11}=S_{22}=-S_{12}=2{\color{black}|e^*I_1|}$, where $e^*$ is the layer-resolved exciton charge. A fractional value of $e^*{\color{black}=e/3}$ would provide evidence of anyonic excitons.

In conclusion, layer-resolved current and noise give ways to probe charge distribution in bilayer excitations. This distribution provides information about their statistics and the topological order.

This research was supported by NSF under Grants DMR-2529089
and PHY-2309135. J.I.A.L. acknowledges support from the Air Force Office of Scientific Research under award no. FA9550-23-1-0482.

\begin{figure}
    \centering
    \includegraphics[width=1.0\linewidth]{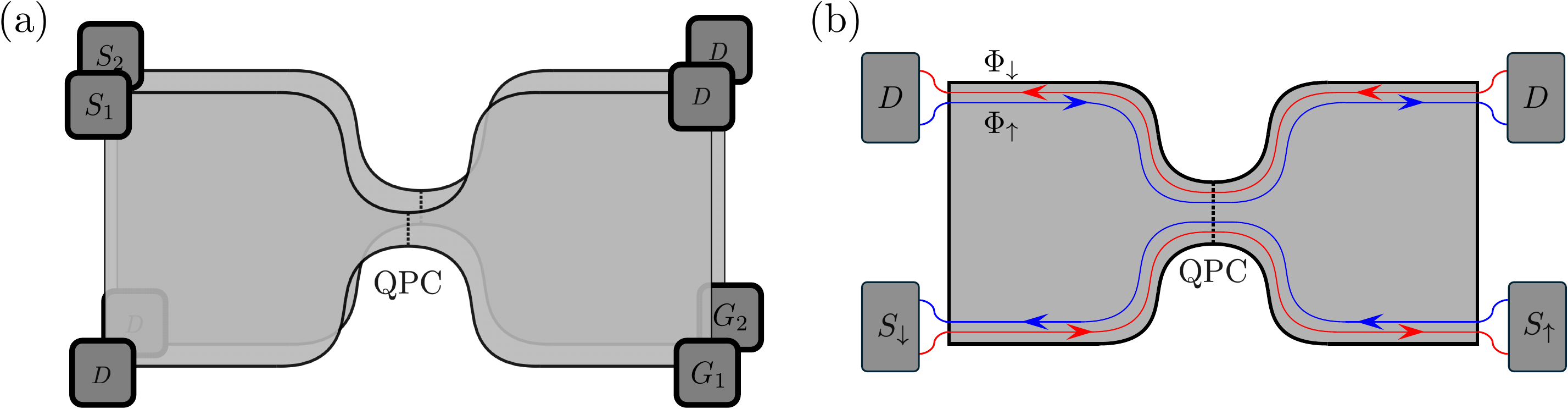}
    \caption{Constriction between two edges. (a) Bilayer. Quasiparticles tunnel across QPC. Layer-resolved sources and gates are placed at the top and bottom edges. (b) Spin Hall effect. The spin channels (red and blue) have opposite chiralities. The four drains and sources control the chemical potentials of the spin-up and -down channels.}
    \label{fig:QPC}
\end{figure}



\begin{figure}
    \centering
    \includegraphics[width=1.0\linewidth]{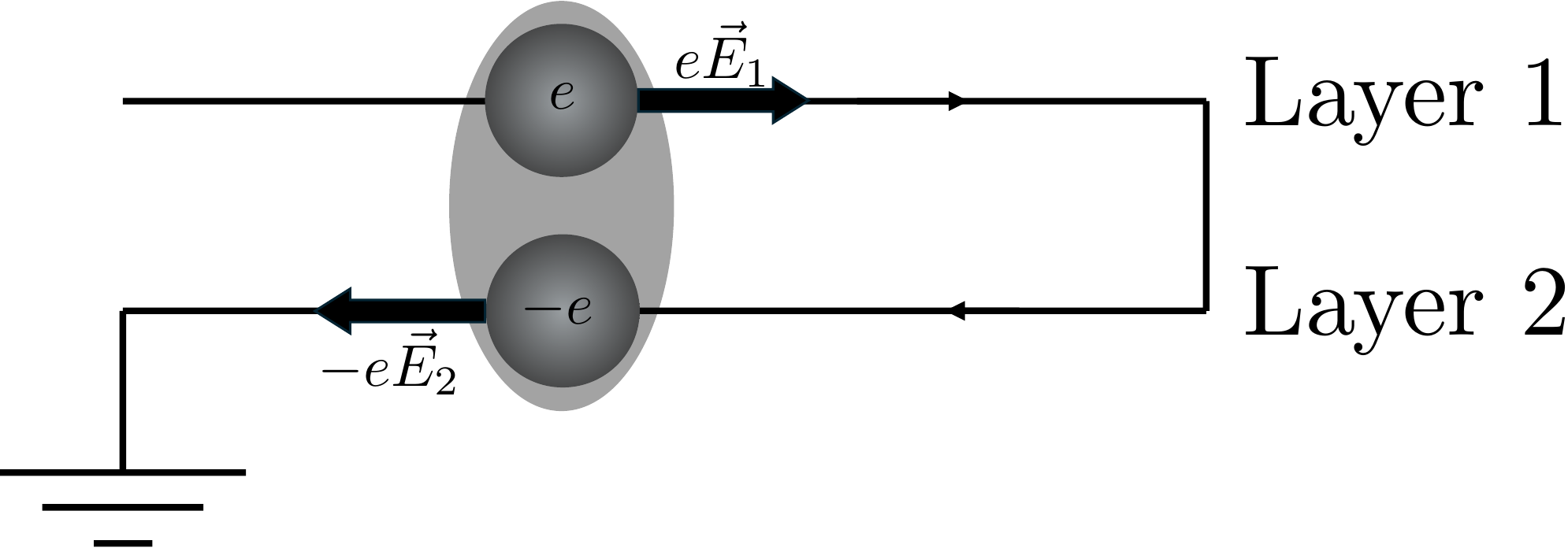}
    \caption{Layer-resolved illustration of a composite fermion Cooper pair in the Pfaffian state. Composite fermions are from the same layer. The pair has a layer-resolved charge $(e,-e)$ and thus experiences a total electric force $e(\vec{E}_1-\vec{E_2})$ with layer-resolved in-plane electric fields $\vec{E}_1,\vec{E}_2$. 
    }
    \label{fig:Pfaffian}
\end{figure}

\makeatletter\auto@bib@empty\let\auto@bib@innerbib\@empty\let\write@bibliographystyle\relax\makeatother

\begin{center}
{\bf End Matter}
\end{center}

\begin{figure}
    \centering
    \includegraphics[width=1.0\linewidth]{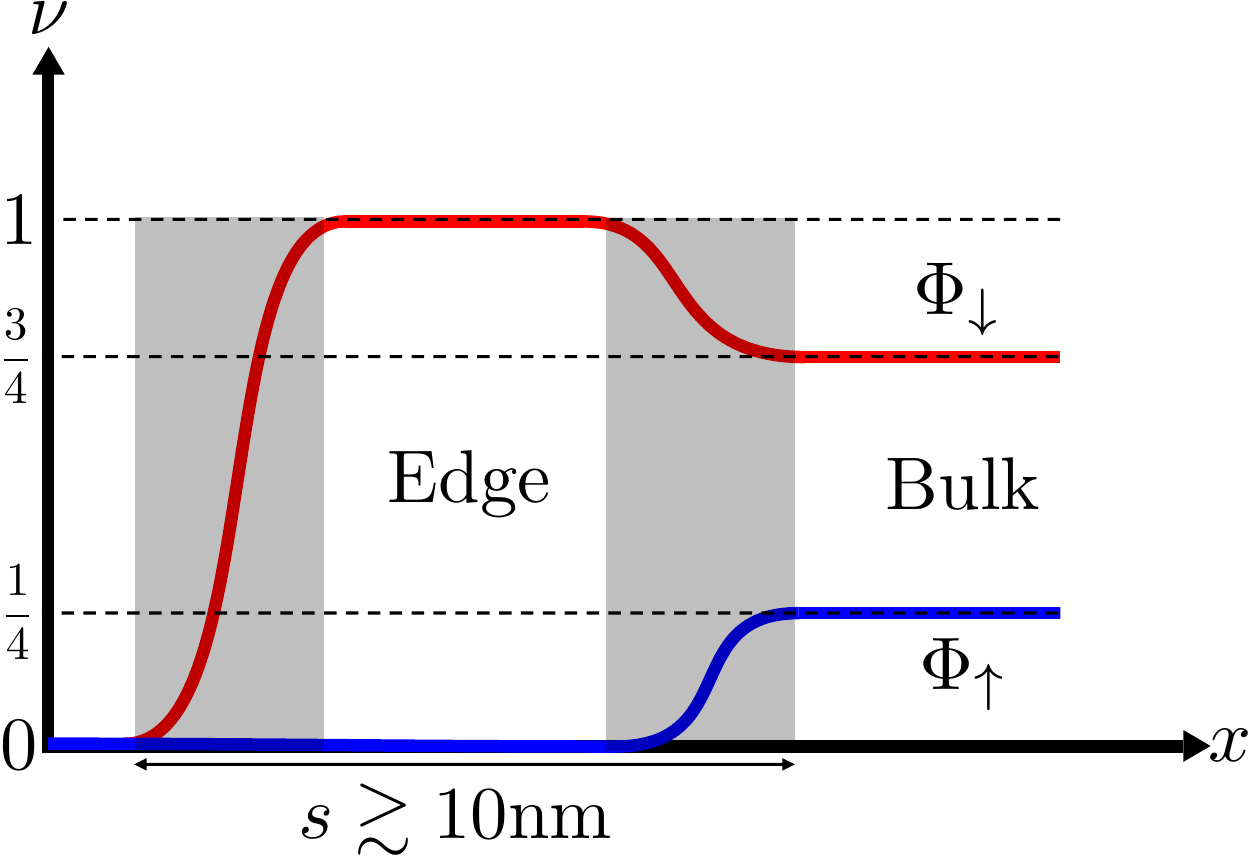}
    \caption{Spin-resolved edge structure in the PH-$mmn$ state. The edge channels are shaded. The outer channel carries spin-down electrons. The inner channels host two modes of opposite spin. 
    }
    \label{fig:spindown}
\end{figure}

{\it Edge theory in the PH-$mmn$ states and additional remarks on the PH-331 state--}
{\color{black} (1) The first part of the End Matter focuses on the edge theory in the PH-$mmn$ states.} To understand transport through a constriction we need to understand the edge theory of the PH-$mmn$ state first. The Lagrangian  includes two contributions. The first is quadratic in the charge densities $(-1)^{k+1}e\partial_x\phi_k/2\pi$ in three channels:

\begin{equation}
L_0=\int dx\sum_{i,j=1}^3\frac{1}{4\pi}(\partial_t\phi_iK_{ij}\partial_x\phi_j-
\partial_x\phi_iU_{ij}\partial_x\phi_j).
\end{equation}
The second contribution comes from random tunneling between two spin-down channels:

\begin{equation}
{\color{black}L_t=\int dx [\xi(x)\exp(in\phi_1+im\phi_2{\color{black}-}i\phi_3)+{\rm h.~c.}],}
\end{equation}
where the tunneling amplitude $\xi(x)$ is a random function of the coordinate.

On a long edge,  tunneling equilibrates the chemical potentials of the spin-down channels. However, in the calculation of the low-voltage current through a constriction, the tunneling contribution can be ignored since it is irrelevant in the renormalization group sense. To see that, it is convenient to change variables into $\phi_c=(\phi_1-\phi_2)$ and 
$\phi_n=\sqrt{n+1}(\phi_1+\phi_2)$. The $K$ matrix simplifies to

\begin{equation}
K=\begin{pmatrix}
1 & 0 & 0 \\
0 & 1 & 0 \\
0 & 0 & -1
\end{pmatrix},
\end{equation}
while the tunneling operator becomes $\exp(i{\color{black}n}\phi_1+i{\color{black}m}\phi_2{\color{black}-}i\phi_3)=
\exp(i\sqrt{n+1}\phi_n-i\phi_c{\color{black}-}i\phi_3)$. To find the scaling dimension of the operator, one needs to diagonalize the interaction matrix $U_{ij}$ with a transformation that preserves the above diagonal form of the $K$-matrix. This is possible since $U_{ij}$ is positive definite in a stable system. The diagonalization procedure turns the tunneling operator into $\exp(i\sum_{k=1}^3a_k\tilde\phi_k)$, where $\tilde\phi_k$ are obtained from the original fields by a $O(2,1)$ pseudorotation. The scaling dimension of the tunneling operator equals $\Delta_n=\sum_{k=1}^3a_k^2/2$.
The pseudorotation preserves the norm of the vector $(a_1,a_2,a_3)$: $a_1^2+a_2^2-a_3^2=(n+1)+1-1=n+1$. It follows that $\Delta=\sum a_k^2/2\ge (a_1^2+a_2^2-a_3^2)/2= (n+1)/2$. It is known \cite{KF1997} that random tunneling is only relevant if its scaling dimension $\Delta_n<3/2$. Thus, our calculation establishes that for all $n>1$, tunneling is irrelevant.

The tunneling operator may, in principle, be relevant at $n=1$, but we do not expect this to be the case in experimental systems. Indeed, screening layers {\color{black} are separated from the 2D electron gas in tMoTe$_2$ devices by a few nanometers to a few tens of nanometers of insulator}. This is {\color{black} comparable to} the  lattice constant $a\approx 10$ nm of the moire lattice. The two spin-down channels of the PH-$mmn$ state are spatially separated (Fig. 3) at the distance $s{\color{black}\gtrsim a} {\color{black}\sim} b$, where $b$ is the distance from the screening layer. We can thus expect the interaction of the counter-propagating modes to be {\color{black}suppressed} and the scaling dimension $\Delta$ to be approximately the same as without interaction. Indeed, the interaction of co-propagating modes has no effect on scaling dimensions \cite{WenBook}. This yields $\Delta{\color{black}\approx}2>3/2$ at $n=1$.

{\color{black}We can now estimate the scaling dimension of the tunneling operator of an anyon described by the vector $t=(a,b,0)$. From the standard prescription \cite{WenBook}, we get the scaling dimension

\begin{equation}
\label{new-2}
g=\frac{m(a^2+b^2)-2nab}{m^2-n^2}.
\end{equation}}

{\color{black} Neither experiment \cite{FSHE,time-reversal} nor theory \cite{mote-numerics} proves
exact spin conservation. In fact, 
scattering off impurities may flip spin. Fortunately, as discussed in the {\color{black}Supplemental Material~\cite{SM}\nocite{StewartGalois,Hall2015,comment}}, spin-flip processes are irrelevant in the low-energy limit {\color{black}and much weaker than spin-preserving inter-channel tunneling}.}

{\color{black}(2) The second part of the End Matter addresses an alternative way to probe the PH-331 state.}
To distinguish it from the states of the 16-fold way one can go to very low temperatures and voltages so that the tunneling of the ${\color{black}(1,0,0)}$ and ${\color{black}(0,1,0)}$ particles dominates. If one applies the layer resolved voltages $V_{\color{black}\downarrow}=-3V_{\color{black}\uparrow}$, only ${\color{black}(0,1,0)}$ particles tunnel. We find that {\color{black}$I_\downarrow=3I_\uparrow$ and $S_{\downarrow\downarrow}=9S_{\uparrow\uparrow}={\color{black}-3S_{\uparrow\downarrow}}=|3eI_\downarrow / {\color{black}4}|$}. A subtlety involves the itinerant nature of the ${\color{black}(1,-1,0)}$-anyons in the PH-$mmn$ state \cite{sodemann}. These particles may have a large localization length, yet, the above equation assumes that no non-tunneling transport of itinerant charged ${\color{black}(1,-1,0)}$-particles is possible. If it happens, the tunneling currents $I_{\uparrow,\downarrow}$ combine with the layer-independent bulk current $I_{{\color{black}(1,-1,0)}}(V=[V_\uparrow+V_\downarrow]/2)$. Note that itinerant particles only couple to the sum of the layer-resolved voltages. To exclude the effect of itinerant anyons,
one should compare layer resolved currents for $V_\uparrow=3V/2,~ V_\downarrow=-V/2$ and $V_\uparrow=-V/2, ~ V_\downarrow=3V/2$. The $I_{{\color{black}(1,-1,0)}}$ contributions to the layer-resolved currents are the same in both cases. In the first case,
$(I_\uparrow - I_{{\color{black}(1,-1,0)}})=3(I_\downarrow-I_{{\color{black}(1,-1,0)}})$. In the second case,
$(I_\uparrow - I_{{\color{black}(1,-1,0)}})=(I_\downarrow-I_{{\color{black}(1,-1,0)}})/3$.

\clearpage
\onecolumngrid
\makeatletter
\input{size11.clo}
\setlength{\textwidth}{\dimexpr\paperwidth-51.2mm\relax}
\setlength{\textheight}{\dimexpr\paperheight-51.2mm\relax}
\setlength{\oddsidemargin}{\dimexpr25.6mm-1in\relax}
\setlength{\evensidemargin}{\oddsidemargin}
\setlength{\topmargin}{\dimexpr25.6mm-1in-\headheight-\headsep\relax}
\setlength{\parindent}{17pt}
\setlength{\parskip}{0pt plus 1pt}
\set@column@hsize\@ne
\set@colht
\def\ps@supplement{%
  \let\@oddhead\@empty\let\@evenhead\@empty
  \def\@oddfoot{\normalfont\hfil\thepage\hfil}%
  \let\@evenfoot\@oddfoot}
\pagestyle{supplement}
\thispagestyle{supplement}
\def\section{\@startsection{section}{1}{\z@}%
  {-3.5ex \@plus -1ex \@minus -.2ex}{2.3ex \@plus .2ex}%
  {\normalfont\Large\bfseries}}
\def\subsection{\@startsection{subsection}{2}{\z@}%
  {-3.25ex\@plus -1ex \@minus -.2ex}{1.5ex \@plus .2ex}%
  {\normalfont\large\bfseries}}
\def\@seccntformat#1{\csname the#1\endcsname\quad}
\let\@sectioncntformat\@seccntformat
\let\@subsectioncntformat\@seccntformat
\let\@hangfrom@section\@undefined
\let\@hangfrom@subsection\@undefined
\let\@hangfroms@section\@undefined
\let\@hangfroms@subsection\@undefined
\def\p@section{}\def\p@subsection{}
\def\thesection{\arabic{section}}
\def\thesubsection{\thesection.\arabic{subsection}}
\def\thefigure{\arabic{figure}}
\def\thetable{\arabic{table}}
\def\theequation{\arabic{equation}}
\def\figurename{Figure}\def\tablename{Table}
\long\def\@makecaption#1#2{%
  \vskip\abovecaptionskip
  \sbox\@tempboxa{#1: #2}%
  \ifdim\wd\@tempboxa>\hsize #1: #2\par
  \else\global\@minipagefalse\hb@xt@\hsize{\hfil\box\@tempboxa\hfil}\fi
  \vskip\belowcaptionskip}
\setlength{\abovecaptionskip}{10pt}
\setlength{\belowcaptionskip}{0pt}
\def\bibsection{\section*{References}}
\def\bibfont{\normalfont\normalsize}
\let\thebibliography\NAT@thebibliography
\let\endthebibliography\endNAT@thebibliography
\let\@bibsetup\NATx@bibsetnum
\def\bibnumfmt#1{[#1]}
\def\NAT@sort{0}\def\NAT@cmprs{0}
\setlength{\bibsep}{4.5pt plus 2pt minus 1pt}
\makeatother
\setcounter{page}{1}
\setcounter{section}{0}
\setcounter{subsection}{0}
\setcounter{equation}{0}
\setcounter{figure}{0}
\setcounter{table}{0}
\setcounter{footnote}{0}
\setcounter{secnumdepth}{3}
\raggedbottom
\def\[{\begin{equation}\begin{aligned}}
\def\]{\end{aligned}\end{equation}}
\def\beq{\begin{equation}}
\def\eeq{\end{equation}}

\vspace*{2em}
\begin{center}
{\LARGE Supplemental Material for ``Probing bilayer topological order with layer-resolved transport''\par}
\vskip 1.5em
{\large Hongquan Liu$^{1,2}$, J.I.A. Li$^{3}$, and D. E. Feldman$^{1,2}$\par}
\vskip 1em
{\large \today\par}
\end{center}
\vskip 1.5em

\begin{center}
$^{1}$Department of Physics, Brown University, Providence, RI, USA\\
$^2$Brown Theoretical Physics Center, Brown University, Providence, Rhode Island 02912, USA\\
$^{3}$Department of Physics, University of Texas at Austin, Austin, TX 78712, USA
\end{center}

\section{Review of topological orders}

\subsection{331 state}
The 331 state (for a review, see Ref. \cite{supp:review-FH}) is an Abelian state characterized by the $K$-matrix
and the charge vector

\[
K = \begin{pmatrix}
    3&1\\
    1&3
\end{pmatrix}, \quad t = (1,1)
\]
so that the filling factor $\nu = t K^{-1} t^T = 1/2$. The ground state wavefunction is given by
\[
\Psi_{331}(\{z_i\},\{w_i\})\sim \prod_{i<j}(z_i-z_j)^3 (w_i-w_j)^3 \prod_{i,j}(z_i-w_j) e^{-\sum_i\left|z_i\right|^2 /4 \ell_B^2-\sum_j\left|w_j\right|^2 /4 \ell_B^2},
\]
where $z_k=x^1_k+iy^1_k$ and $w_k=x^2_k+iy^2_k$ are the complex coordinates of the electrons in the two layers.

On the edge, we have two co-propagating modes with the charge densities $\frac{e\partial_x\phi_k}{2\pi}$, $k=1,2$. The edge Lagrangian

\begin{equation}
L=\int dx \sum_{i,j=1}^2\frac{1}{4\pi}(\partial_t\phi_iK_{ij}\partial_x\phi_j-
\partial_x\phi_iU_{ij}\partial_x\phi_j),
\end{equation}
where $U_{ij}$
 describes interactions.
  The two most relevant electron operators on the edge are $\Psi_e = \exp(i l\cdot  \boldsymbol{\phi})$ where $l = (3,1)$ or $(1,3)$ and $\boldsymbol{\phi}=(\phi_1,\phi_2)$. Quasiparticles are
  created by the operators $\exp(i l\cdot  \boldsymbol{\phi})$
  with  integral $l$-vectors, and the lowest-charge excitations  are described by $l=(1,0)$ or $(0,1)$. The most relevant neutral exciton is described by $l=(1,-1)$. The component-wise charges and the tunneling scaling dimensions $g=2\Delta$ of the quasiparticle operators are listed in Table~\ref{supp:tab:331}.

\begin{table}[t]
\centering
\begin{tabular}{cccc}
\hline\hline
Quasiparticle & $l$-vector &  $(q_1,q_2)$ & $g$ \\
\hline
Elementary Anyon & $(1,0)$, $(0,1)$ & $(3e/8,-e/8)$, $(-e/8,3e/8)$& $3/8$ \\
Exciton & $(1,-1)$ & $(e/2,-e/2)$ & $1$\\
Electron & $(3,1)$, $(1,3)$ & $(e,0)$, $(0,e)$ & $3$ \\
\hline\hline
\end{tabular}
\caption{Quasiparticles in the 331 state. }
\label{supp:tab:331}
\end{table}

\subsection{Pfaffian state}
The Pfaffian state (for a review, see Ref. \cite{supp:review-FH}) is a non-Abelian state characterized by a one-dimensional $K$-matrix $K=2$, one-dimensional charge vector $t=1$, and an Ising sector. Its edge theory has the Lagrangian density 
\[
 \mathcal{L} = 
 i\psi(\partial_t+v_n\partial_x)\psi-\frac{2}{4\pi}\partial_x\phi(\partial_t+v_c\partial_x)\phi,
\]
where $\psi$ is a neutral Majorana fermion with the speed $v_n$, the charge density is $\frac{e\partial_x\phi}{2\pi}$, and $v_c$ is the speed of the charged mode.
The electron operator on the edge is given by $\Psi_e = \psi e^{2i\phi}$. Quasiparticles are created by operators $O e^{ia\phi}$ where $O=1,\psi$ if $a$ is an integer and $O=\sigma$ if $a$ is a half-integer. We summarize  quasiparticle properties in Table~\ref{supp:tab:Pfaffian}.

The ground state wavefunction in a single layer is given by 
\[
\label{supp:Sup-5}
\Psi_{\mathrm{Pfaffian}}(\{z_i\}) \sim
\mathrm{Pf}\left[\frac{1}{z_i-z_j}\right]\cdot \prod_{i<j} (z_i-z_j)^2 
e^{-\sum_i\left|z_i\right|^2 /4 \ell_B^2},
\]
where ${\mathrm{Pf}}$ denotes the Pfaffian of an antisymmetric matrix, and $z_k=x_k+iy_k$ are electron coordinates. 
This wave function can be understood as the ground state of $p+ip$-paired composite fermions. This is because a $p+ip$ BCS wavefunction has the form of $\mathrm{Pf}[1/(z_i-z_j)]$ and the Jastrow factor $(z_i-z_j)^2$ implies the attachment of two flux quanta to each electron. The wave function can be generalized to a bilayer by assigning to each electron a layer index 1 or 2 and substituting $z^1_j$ in place of half of the coordinates $z_j$ and $z^2_j$ in place of the other half of the coordinates $z_j$ in equation (\ref{supp:Sup-5}).
The wave function should also be antisymmetrized over all the choices of subscripts $j$, which are paired with superscripts 1 and 2.


\begin{table}[t]
\centering
\begin{tabular}{ccc}
\hline\hline
Quasiparticle & Operator  & $g$ \\
\hline
$e/4$-Anyon & $\sigma e^{i\phi/2}$ & $1/4$ \\
$e/2$-Anyon & $e^{i\phi}$ & $1/2$\\
Electron & $\psi e^{2i\phi}$ &   $3$ \\
\hline\hline
\end{tabular}
\caption{Quasiparticles in the Pfaffian state. }
\label{supp:tab:Pfaffian}
\end{table}

\subsection{PH-{$mmn$} states}
The PH-$mmn$ state \cite{supp:sodemann} is an Abelian state proposed to describe the $\nu=3$ quantum spin Hall effect in tMoTe$_2$. Electrons fill two bands of opposite spin and opposite Chern numbers. The fully filled bands contribute two quanta of the spin Hall conductance. Two higher bands of opposite spin and opposite Chern numbers are partially filled with the total filling factor of $\nu_f=1$. This can be interpreted as an equal number of electrons of one spin polarization and holes of the other polarization.

If we assign coordinates $z_k=x_k+iy_k$ and $w_q=x_q+iy_q$ to the electrons of one polarization and the holes of the other polarization respectively, the wave function assumes the form of the Halperin $mmn$ state,

\[
\Psi_{\rm  PH}(\{z_i\},\{w_i\})\sim \prod_{i<j}(z_i-z_j)^m (w_i-w_j)^m \prod_{i,j}(z_i-w_j)^n e^{-\sum_i\left|z_i\right|^2 /4 \ell_B^2-\sum_j\left|w_j\right|^2 /4 \ell_B^2}.
\]
This corresponds to the $K$-matrix and the spin-resolved charge vectors of the form

\[
K_{mmn} = \begin{pmatrix}
    m&n\\
    n&m
\end{pmatrix}, \quad t_\uparrow=(1,0), \quad t_\downarrow = (0,1).
\]
To account for the particle hole-transformation, we need to add a column and a row to the $K$-matrix so that
\[
K = \begin{pmatrix}
    m&n&0\\
    n&m&0\\
    0&0&-1
\end{pmatrix}, \quad t_\uparrow=(1,0,0), \quad t_\downarrow = (0,-1,1).
\]
On the edge we have two right-moving modes $\phi_{1,2}$ and one left-moving mode $\phi_3$ as discussed in the main text. Quasiparticle operators on the edge are defined in terms of integral vectors $l=(l_1,l_2,l_3)$
and can be expressed as $\exp(i\sum_{k=1}^3l_k\phi_k)$.

Spin-up and -down electrons correspond to $l_\uparrow = (m,n,0)$ and $l_\downarrow = (-n,-m,0)$. The elementary anyons correspond to $l_\uparrow=(1,0,0)$ and $l_\downarrow=(0,1,0)$. Notably, an exciton described by $l=(1,1,0)$ is created by the most relevant operator in the edge theory when $n\ge 3$. 

A special property \cite{supp:sodemann} of this PH-$mmn$ state is that it contains an itinerant charged quasiparticle $(1,-1,0)$, We summarize the properties of quasiparticles in the PH-$mmn$ state in Table~\ref{supp:tab:PHmmn}.

\begin{table}[t]
\centering
\begin{tabular}{cccc}
\hline\hline
Quasiparticle & $l$-vector &  $(q_1,q_2)$ & $g$ \\
\hline
Elementary Anyon & $(1,0,0)$, $(0,1,0)$ & $(3e/8,e/8)$, $(e/8,3e/8)$& $3/8$ \\
Exciton & $(1,1,0)$ & $(e/2,-e/2)$ & $1/(n+1)$\\
Itinerant Anyon & $(1,-1,0)$ & $(e/2,e/2)$ & {\color{black}$1$}\\
Electron & $(m,n,0)$, $(-n,-m,0)$ & $(e,0)$, $(0,e)$ & $3$ \\
\hline\hline
\end{tabular}
\caption{Quasiparticles in the PH-$mmn$ state. }
\label{supp:tab:PHmmn}
\end{table}

\subsection{Bilayer states of the 16-fold way}
Such states were considered in Ref. \cite{supp:fsqh-3}. Each state should be understood as a pair of fractional quantum Hall liquids of the 16-fold way (for a review, see \cite{supp:review-FH}) in two bands of the opposite spin polarization on top of the two filled bands, corresponding to $\nu=2$.
The simplest example would be two Pfaffian liquids with the edges of the opposite chirality.
Ref. \cite{supp:fsqh-3} assumes that the two liquids are connected by the time-reversal symmetry, but this property makes no difference for this paper. The only important property for us is the quasiparticle charge. Each quasiparticle can be understood as a combination of two quasiparticles in the two spin-polarized topological liquids. The minimal charge of a quasiparticle in each liquid is $e/4$. Such a particle can be combined with a neutral excitation in the other liquid so 
that the minimal overall charge is $e/4$.


\subsection{Minimal fractional topological insulator}

This state was introduced in Ref. \cite{supp:fsqh-2}. It can be obtained from a pair of Pfaffian liquids with the edges of the opposite chirality by condensing pairs of neutral Majorana fermions from the two liquids (see Section 1.2). The condensation confines anyons of charge $e/4$, and the minimal nonzero excitation charge becomes $2\times e/4=e/2$. In the limit of low voltages and temperatures, the most relevant  quasiparticle operator creates a pair of charges $e/4$ in the two spin-polarized bands.

The edge theory contains only two contra-propagating Bose channels since the condensation of Majorana fermions gaps them out. Since an edge electron operator in the Pfaffian state contains a Majorana fermion, electrons are gapped on the edges of the minimal fractional topological insulator. Thus, only electron pairs can tunnel into the edge from a conductor.

\subsection{Bilayer states in graphene}

Multiple experimentally relevant states in bilayer graphene are discussed in Refs. \cite{supp:Zhang2025fractionalexciton,supp:3/8+3/8}. We give only one example here, relevant for the experimental data from Ref. \cite{supp:3/8+3/8}. Consider a bilayer with two Laughlin liquids at the filling factor $1/3$ in each layer. Ref. \cite{supp:3/8+3/8} suggests that when slightly away from  the center of the plateau, transport is dominated by composite neutral quasiparticles, which bind an anyon of charge $e/3$ in one layer and an anyon of charge $-e/3$ in the other layer. The cross-correlation noise is a natural technique to test this prediction.

\section{Cooper pairs in the bilayer Pfaffian state}
As discussed in Section 1.2, the Pfaffian liquid can be understood as a superconductor of composite fermions. A composite fermion is electrically neutral. Indeed, it is made of an electron of charge $e$ and two flux quanta. Inserting a flux quantum creates an effective electric field that drives a current in a quantum Hall liquid. One can easily compute it as $-n\nu e$, where $\nu$ is the filling factor and $n$ is the number of the flux quanta attached to an electron. At $n=2$ and $\nu=1/2$, we get a neutral excitation. This is, of course, well known for single-component systems. There is a twist when we have two components. Indeed, the charge of an electron must be localized in one component. The flux, on the other hand, equally affects both components. Thus, a composite fermion represents  a dipole in the component space with the component charges $\pm e/2$. 

\begin{figure}
    \centering
    \includegraphics[width=0.5\linewidth]{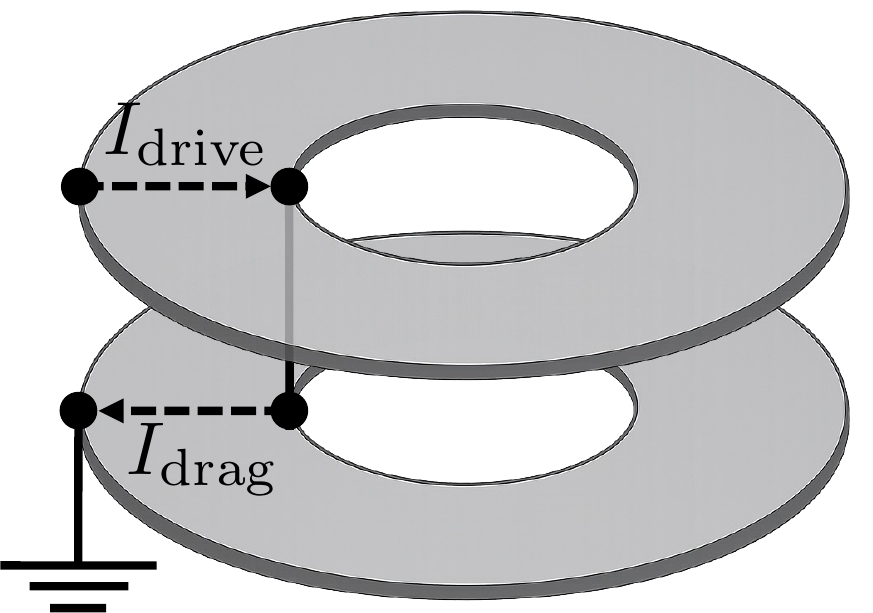}
    \caption{Schematics of a counter-flow experiment in a quantum Hall bilayer sample with the Corbino geometry.}
    \label{supp:fig:Corbino-first}
\end{figure}

\begin{figure}
    \centering
    \includegraphics[width=0.5\linewidth]{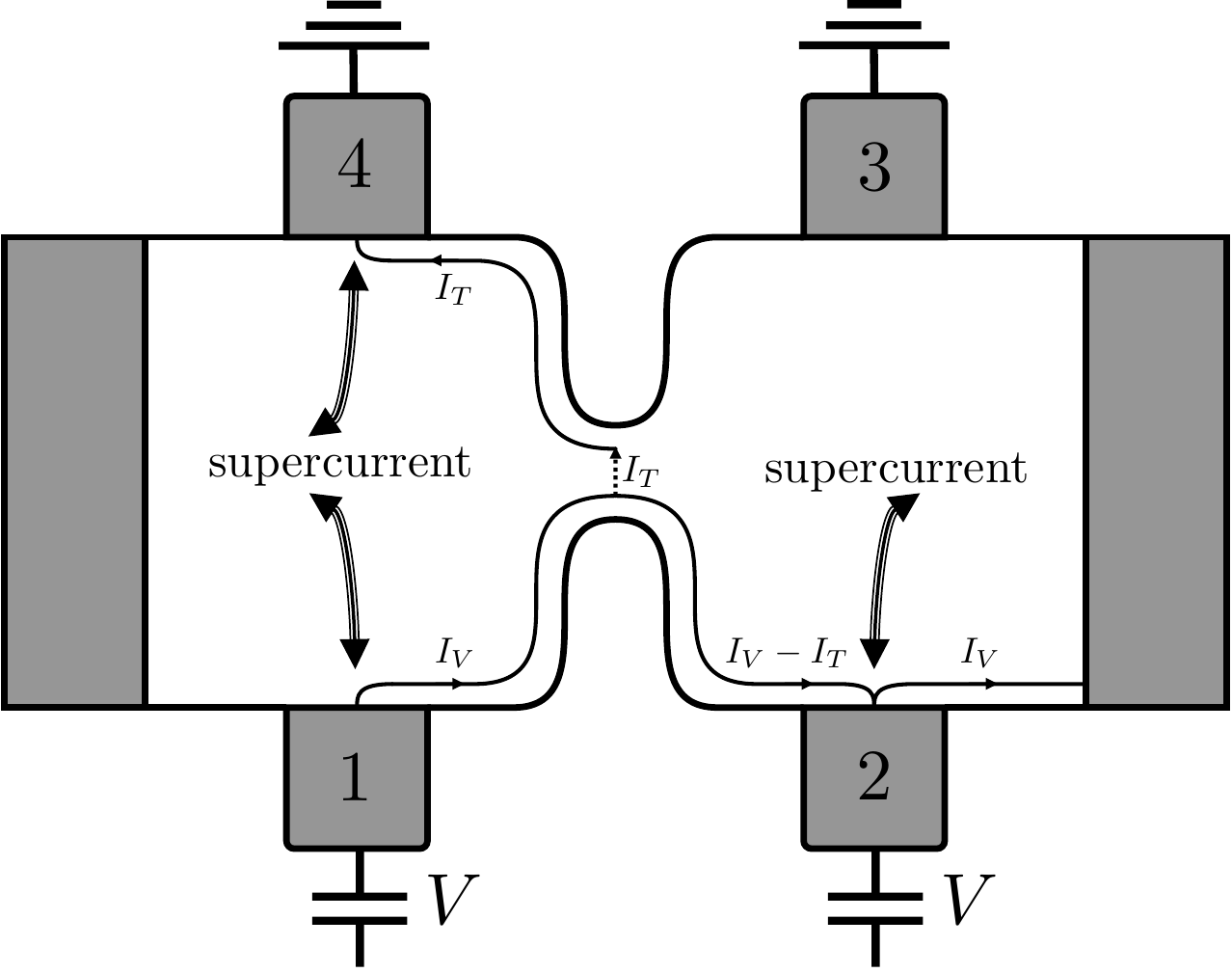}
    \caption{Schematics of a tunneling experiment in a bilayer Pfaffian liquid.  Electrical contacts on the left and on the right are connected to both layers simultaneously, whereas contacts 1 through 4 are layer specific: contacts 1 and 2 (3 and 4) connect to lower (upper) layer only. 
    }
    \label{supp:fig:Corbino}
\end{figure}

A Cooper pair made of two electrons from the same component is also a dipole in the component space with the component charges $\pm e$. The condensate of such Cooper pairs is a superfluid. Hence, the quantum Hall liquid rearranges its charges in a way that eliminates the total electric force on the Cooper pair. In other words, the electric fields acting on the two components must be identical. The existence of inter-component Cooper pairs does not impose constraints since then the charge of each component, as well as overall  charge across two layers, is zero. 

One important consequence of interlayer Cooper pairing is the counterflow effect, which has been observed across a wide range of material platforms~\cite{supp:Zhang2025fractionalexciton,supp:3/8+3/8,supp:Kellogg2002exciton,supp:Kellogg2004exciton,supp:Tutuc2004exciton,supp:Nandi2012exciton,supp:Li2017exciton,supp:Liu2017exciton}. In quantum Hall graphene bilayers, counterflow measurements performed in the Corbino geometry have been used to identify interlayer Cooper pairs, including those with potentially fractional statistics. The measurement configuration is illustrated in Fig.~1. A current is injected into the upper layer and extracted from the lower layer, while the two layers are electrically connected along one edge of the Corbino disk. In the presence of a charge gap, neutral interlayer Cooper pairs can carry a dissipationless supercurrent through the bulk. This exciton supercurrent generates equal-magnitude electrical currents flowing in opposite directions in the two layers, giving rise to the observed counterflow response. The Corbino geometry is particularly advantageous because it suppresses edge-state transport, allowing the bulk transport of interlayer Cooper pairs to be probed directly.

What happens if one attempts to measure the layer-resolved current and noise in a Hall bar geometry (Fig. 2)? Electrical contacts on the left and on the right are connected to both layers simultaneously, whereas contacts 1 through 4 are layer specific. Contacts 1 and 2 connect to the lower layer only, and contacts 3 and 4 connect to the upper layer. A tunneling contact is made in the center in the form of a constriction in the sample. The current $I_V=e^2V/2h$ flows from the lower left contact towards the tunneling contact. The tunneling current $I_T$ flows from the tunneling constriction to the upper left contact. The current from the tunneling constriction to the lower right contact equals $I_V-I_T$. The right lower contact is maintained at voltage $V$, so that the current that leaves the contact towards the right side of the device is $I_V$. Thus, the contact absorbs the current $-I_T$. The upper left contact absorbs the current $I_T$.
It follows that the identical currents $|I_T|$ and the identical noises of the current $I_T$ are detected in contacts 2 and 4. Since the charge of the tunneling quasiparticles is $e/4$, the noise $S=e|I_T|/2$.

This may seem puzzling since contact 1 injects charge into the lower layer only while contact 4 absorbs charge from the upper layer only. To understand the transport we need to remember about neutral supercurrents  of inter-layer Cooper pairs in the device. In particular, the supercurrent entering contact 1 from the bulk (Fig. 2) helps equalize the layer-resolved currents that flow from contact 1 towards the tunneling constriction. Crucially, neutral supercurrents do not affect the total electric current or noise in the system. Thus, the currents and noises, detected in contacts 2 and 4 are exactly the same as on the edge segments between those contacts and the tunneling constriction. At the same time, supercurrents help transform layer-unresolved currents and noises into layer-resolved ones.

\section{Spin-flip processes}

In this section, we show that spin-flip processes on the edge {\color{black}of a PH-$mmn$ liquid} are irrelevant in the renormalization group sense at low energies.

Any spin-flip operator in the Hamiltonian must conserve charge and braid trivially with every quasiparticle. In addition, it must be a 
Bose-operator, but that property does not add any new restrictions on the operator form. The allowed operators $\exp(a\phi_1+b\phi_2+c\phi_3)$ are constrained by the conditions

\begin{equation}
a=2k(n+1)-nc;
\end{equation}
\begin{equation}
b=2k(n+1)-(n+2)c
\end{equation}
with a nonzero integer $k$.
The scaling dimension of the spin-flip operator equals

\begin{equation}
\Delta_s=c^2+\frac{(2k-c)^2(n+1)}{2}.
\end{equation}
Due to the momentum mismatch between the modes, we are not interested in nonrandom tunneling \cite{supp:comment}. For disorder-mediated tunneling, it is sufficient to verify that the scaling dimension $\Delta_s>3/2$. This is the case for $|c|>1$ since then $\Delta_s\ge c^2\ge 4$. This is the case at $c=0$ since then $\Delta_s=2k^2(n+1)\ge 4$. Finally, at $|c|=1$,
$\Delta_s\ge c^2+(n+1)/2\ge 2$.

\section{Scaling dimensions with interactions}
Starting with the edge theory for the PH-$mmn$ state in MoTe$_2$,
\[
\mathcal{L}_0= \sum_{i, j=1}^3\frac{1}{4\pi}\left(\partial_t \phi_i K_{i j} \partial_x \phi_j-\partial_x \phi_i U_{i j} \partial_x \phi_j\right),
\]
we perform a change of variables $(\phi_1,\phi_2,\phi_3)\mapsto (\phi_n,\phi_c,\phi_3)$, where $\phi_n = \sqrt{n+1}(\phi_1+\phi_2)$ and $\phi_c =\phi_1-\phi_2$.
The $K$ matrix becomes a (2+1)D Minkowski metric
\[
\label{supp:appendix-2}
K'=\eta = \begin{pmatrix}
1 & 0 & 0\\
0 & 1 & 0\\
0&0&-1
\end{pmatrix}.
\]
This is equivalent to applying the following linear transformation to the basis $(\phi_1,\phi_2,\phi_3)$:
\[
\begin{pmatrix}
\phi_1\\
\phi_2 \\
\phi_3
\end{pmatrix}
=
W
\begin{pmatrix}
\phi_n\\
\phi_c \\
\phi_3
\end{pmatrix}, 
\quad
W = \left(
\begin{array}{ccc}
 \dfrac{1}{2\sqrt{n+1}} & \dfrac{1}{2} & 0 \\
 \dfrac{1}{2\sqrt{n+1}} & -\dfrac{1}{2} & 0 \\
 0 & 0 & 1 \\
\end{array}
\right).
\]
Therefore we can rewrite our edge theory as
\[
\mathcal{L}_0=
\frac{1}{4\pi}\sum_{i,j=1}^3\left[
\partial_t \phi'_i \eta_{i j} \partial_x \phi'_j 
- \partial_x \phi'_i (W^TUW)_{ij}\partial_x \phi'_j\right], \quad \phi'=(\phi_n,\phi_c,\phi_3).
\]

Since $U$ is a positive definite matrix, we can diagonalize it simultaneously with $K'$ so that the form of
equation (\ref{supp:appendix-2}) is preserved. {\color{black}Any transformation $\Lambda$ that preserves $K'$ is
 an O(2,1) rotation, but it is enough to consider SO(2,1) because a sign difference does not influence the scaling dimension.} We introduce a new basis $\varphi_i$ according to
\[
\begin{pmatrix}
\phi_n\\
\phi_c \\
\phi_3
\end{pmatrix}
=
\Lambda
\begin{pmatrix}
\varphi_1\\
\varphi_2 \\
\varphi_3
\end{pmatrix}.
\]
Any element of SO(2,1) can be decomposed into a product of a boost and two rotations by merit of the $KAK$-decomposition \cite{supp:Hall2015}: $\Lambda =R'_3 B R_3$ where $B$ is a boost and $R_3$ and $R'_3$ are rotations within the spacelike subspace. 
The boost matrix $B$ is similar to a boost in the subspace, spanning the time-like direction and the first space-like direction:

\[
B = R_3(a)B_2(b) R_3(-a),
\]
where $R_3$ is a rotation matrix within the spacelike subspace and $B_2$ is a boost:
\[
R_3(a) = 
\begin{pmatrix}
\cos a & -\sin a & 0\\
\sin a & \cos a & 0\\
0&0&1
\end{pmatrix},\quad 
B_2(b) = 
\begin{pmatrix}
\cosh b & 0 & \sinh b\\
0 & 1 & 0\\
\sinh b&0&\cosh b
\end{pmatrix}.
\]
Therefore $\Lambda \in $ SO(2,1) can be parametrized by
\[
\Lambda = R_3(a)B_2(b) R_3(-a)R_3(c) \equiv R_3(a)B_2(b) R_3(\theta). 
\]
The interaction matrix in the new basis becomes diagonal,
$W^TUW\mapsto \Lambda^T W^T U W \Lambda$. Our edge theory can now be written as
\[
\mathcal{L}_0=
\frac{1}{4\pi}\sum_{i,j=1}^3\left[
\eta_{i j}\partial_t \varphi_i  \partial_x \varphi_j 
-
 (v_i\delta_{ij})\partial_x \varphi_i \partial_x \varphi_j\right].
\]
This allows us to compute the scaling dimensions of different operators.

To calculate the scaling dimension $g_t/2$ of the quasiparticle operator $\exp(i\phi_t)=\exp(t^T\phi)$,
where $\phi^T=(\phi_1,\phi_2,\phi_3)$,
one must switch to the new basis. From
$\phi_t = t^T\phi = t^T W \phi' = t^T W \Lambda \varphi$ we see that in the new basis, the vector $t$ is transformed to
$
t'=(W\Lambda)^T t
$
such that $\phi_t = t'^T \varphi$.
Thus, the scaling dimension is computed from
\[
g_t = t'^T |\eta^{-1}| t' = t^T (W\Lambda)I(W\Lambda)^T t = t^T (W\Lambda)(W\Lambda)^T t.
\]
Here, the rotation $R_3(\theta)$ does not affect the scaling dimension so we can set $\theta=0$. This is because
\[
 (W\Lambda)(W\Lambda)^T 
=& WR_3(a)B_2(b)R_3(\theta) R_3^T(\theta) B_2^T(b) R_3^T(a) W^T\\ 
=& WR_3(a)B_2(b) B_2^T(b) R_3^T(a) W^T\\
=&WR_3(a)B_2(2b) R_3(-a) W^T.
\]

When $n=1$, the most relevant tunneling quasiparticles are ${\color{black}(1,0,0)}$ and ${\color{black}(0,1,0)}$. The corresponding tunneling exponents are calculated from
\[
\label{supp:appendix-13}
g_{{\color{black}(1,0,0)}} =& [(W\Lambda)(W\Lambda)^T]_{11};\\
g_{{\color{black}(0,1,0)}} =& [(W\Lambda)(W\Lambda)^T]_{22}.
\]
They are given by
\[
\label{supp:appendix_g_1_0}
g_{{\color{black}(1,0,0)}} =& \frac{(n+2) \cosh
   ^2 b+\left[2 \sqrt{n+1}\sin (2 a) -n \cos (2 a)\right]\sinh ^2 b }{4(n+1)}\\
   =&
   \frac{3 \cosh
   ^2 b+\left[2 \sqrt{2}\sin (2 a) - \cos (2 a)\right]\sinh ^2 b }{8}
   \\
   =&
   \frac{3}{8}\left[1+ 2\sin^2(a+{\color{black}\alpha})\cdot \sinh^2 b\right];
   \]

\[   
\label{supp:appendix-14}
g_{{\color{black}(0,1,0)}} =& \frac{(n+2) \cosh
   ^2 b-\left[2 \sqrt{n+1}\sin (2 a) +n \cos (2 a)\right]\sinh ^2 b }{4(n+1)}
   \\
   =&
   \frac{3 \cosh
   ^2 b-\left[2 \sqrt{2}\sin (2 a) + \cos (2 a)\right]\sinh ^2 b }{8}
   \\
   =&
   \frac{3}{8}\left[1+ 2\sin^2(a-{\color{black}\alpha})\cdot \sinh^2 b\right],
\]
where ${\color{black}\alpha} = \sin^{-1}(1/\sqrt{3})$.
When there is no inter-mode interaction, then $a=b=0$ and $g_{{\color{black}(1,0,0)}}=g_{{\color{black}(0,1,0)}}=3/8$.

When $n>1$, the most relevant quasiparticles are $\pm (1,1,0)$. The tunneling exponent is calculated from
\[
\label{supp:appendix-15}
g_{{\color{black}(1,1,0)}} =& [(W\Lambda)(W\Lambda)^T]_{11}+2[(W\Lambda)(W\Lambda)^T]_{12}+[(W\Lambda)(W\Lambda)^T]_{22}\\
=&\frac{1}{n+1}\left( 1+2\cos^2 a \cdot \sinh^2 b \right).
\]
When there are no intermode interactions, then $a=b=0$ and $g_{{\color{black}(1,1,0)}}=1/(n+1)$.

To connect the  angles $a$ and $b$ with the symmetric interaction matrix $U$, we may express $a$ and $b$ in terms of the elements of the matrix $u_{ij}$,
\[
U = 
\begin{pmatrix}
u_{11} & u_{12} & u_{13}\\
u_{12} & u_{22} & u_{23}\\
u_{13}&u_{23}&u_{33}
\end{pmatrix}.
\]
 From the parametrization of $\Lambda$, we know that
\[
 \Lambda^T W^T U W \Lambda = 
 \begin{pmatrix}
v_1 & 0 & 0\\
0 & v_2 & 0\\
0&0&v_3
\end{pmatrix} = V,
\]
where $v_i>0$. Therefore,
\[
U
= &
{W^{-1}}^T\Lambda^T(a,b,\theta)^{-1}
V
\Lambda(a,b,\theta)^{-1}W^{-1}
\\
=&
{W^{-1}}^T R_3(a)B_2(-b)R_3(\theta)
V
R_3^T(\theta)B_2(-b)R_3^T(a)
W^{-1}.
\]
Define the $Z$-matrix as
$
Z = W^T U W
$
and the $M$-matrix as
$M = \Lambda^T(a,b,\theta)^{-1}=\Lambda(a,-b,\theta)=R_3(a)B_2(-b)R_3(\theta)$
such that
\[
\label{supp:appendix-18}
Z
=&
 M
V
M^T
\]
with
\[
\label{supp:appendix-20}
\begin{pmatrix}
z_{11} & z_{12} & z_{13}\\
z_{12} & z_{22} & z_{23}\\
z_{13} & z_{23} & z_{33}
\end{pmatrix}
=
\left(
\begin{array}{ccc}
 \dfrac{u_{11}+2 u_{12}+u_{22}}{4 n+4} & \dfrac{u_{11}-u_{22}}{4 \sqrt{n+1}} &
   \dfrac{u_{13}+u_{23}}{2 \sqrt{n+1}} \\
 \dfrac{u_{11}-u_{22}}{4 \sqrt{n+1}} & \dfrac{u_{11}-2 u_{12}+u_{22}}{4}  &
   \dfrac{1}{2} \left(u_{13}-u_{23}\right) \\
 \dfrac{u_{13}+u_{23}}{2 \sqrt{n+1}} & \dfrac{1}{2} \left(u_{13}-u_{23}\right) & u_{33} \\
\end{array}
\right)
\]
and
\[
M  = 
\left(
\begin{array}{ccc}
 \cos a \cosh b \cos \theta -\sin a \sin \theta  & -\sin a \cos \theta -\cos a \cosh b \sin \theta  & -\cos a \sinh b \\
 \sin a \cosh b \cos \theta +\cos a \sin \theta  & \cos a \cos \theta -\sin a \cosh b \sin \theta  & -\sin a \sinh b \\
 -\sinh b \cos \theta  & \sinh b \sin \theta  & \cosh b \\
\end{array}
\right).
\]
From equation (\ref{supp:appendix-18}) we know that
\[
Z = M V \eta \eta M^T = M V \eta M^{-1} M \eta M^T  = M V \eta M^{-1}\eta
\]
and so
\[
Z\eta &= M(V\eta)M^{-1};\\
\begin{pmatrix}
z_{11} & z_{12} & -z_{13}\\
z_{12} & z_{22} & -z_{23}\\
z_{13} & z_{23} & -z_{33}
\end{pmatrix} &= M \begin{pmatrix}
v_1 & 0 & 0\\
0 & v_2 & 0\\
0&0&-v_3
\end{pmatrix} M^{-1}.
\]
This indicates that the columns of $M$ are the eigenvectors of the matrix $Z\eta$, and we only need the third column of $M$, i.e., the third eigenvector of $Z\eta$ to obtain $\tan a$ and $|\tanh b|$. The absolute value  $v_3$ of the corresponding eigenvalue is the unique positive solution of the cubic equation $\det (Z\eta + \lambda I)=0$, which expands to
$
\lambda^3 + A\lambda^2 + B\lambda + C = 0$ with
\[
A =& \mathrm{Tr}[Z\eta] = z_{11}+z_{22}-z_{33};
\]
\[
B =& 
 z_{11} z_{22}-z_{11} z_{33}-z_{22} z_{33}-z_{12}^2+z_{13}^2+z_{23}^2;
 \]
 \[
C =& \det[Z\eta] = -\det Z = -z_{11} z_{22} z_{33}+z_{11} z_{23}^2+z_{22} z_{13}^2+z_{33} z_{12}^2-2 z_{12} z_{13} z_{23}.
\]
The solutions of a cubic equation can be conveniently expressed in terms of two parameters $p$ and $q$,

\[
p=&-\frac{A^2}{3}+B;
\]
\[
q=&\frac{2 A^3}{27}-\frac{A B}{3}+C.
\]
We know that our cubic equation has three real solutions, which means that $p<0$ and $(p/3)^3+(q/2)^2<0$.
We want to find the only positive solution, that is, the largest solution $v_3$. This can be done with a trigonometric method \cite{supp:StewartGalois},

\[
v_3=-\frac{A}{3}+2 \sqrt{-\frac{p}{3}} \cos \left[\frac{1}{3} \arccos \left(\frac{3 q}{2 p} \sqrt{-\frac{3}{p}}\right)\right].
\]

Now that we have obtained $v_3$, we may solve for the eigenvector $(x,y,z)^T\sim (x/z,y/z,1)^T$. Thus 
\[
\begin{pmatrix}
z_{11}+v_3 & z_{12} \\
z_{12} & z_{22}+v_3
\end{pmatrix} 
\begin{pmatrix}
x/z \\
y/z
\end{pmatrix}  = \begin{pmatrix}
z_{13} \\
z_{23}
\end{pmatrix}.
\]
Therefore
\[
\label{supp:appendix_31}
\frac{x}{z}=-\cos a \tanh b=\frac{\left(z_{22}+v_3\right) z_{13}-z_{12} z_{23}}{\left(z_{11}+v_3\right)\left(z_{22}+v_3\right)-z_{12}^2};
\]
\[
\label{supp:appendix_32}
\frac{y}{z}=-\sin a \tanh b=
\frac{\left(z_{11}+v_3\right) z_{23}-z_{12} z_{13}}{\left(z_{11}+v_3\right)\left(z_{22}+v_3\right)-z_{12}^2}
\]
and
\[
\tan a = \frac{\left(z_{11}+v_3\right) z_{23}-z_{12} z_{13}}{\left(z_{22}+v_3\right) z_{13}-z_{12} z_{23}};
\]
\[
\label{supp:appendix_tanhb}
|\tanh b|=\frac{\sqrt{\left[\left(z_{22}+v_3\right) z_{13}-z_{12} z_{23}\right]^2+\left[\left(z_{11}+v_3\right) z_{23}-z_{12} z_{13}\right]^2}}{|\left(z_{11}+v_3\right)\left(z_{22}+v_3\right)-z_{12}^2|}.
\]
The above two equations are sufficient to compute the scaling dimensions (\ref{supp:appendix-13}-\ref{supp:appendix-15}).

When we turn off the interaction of counter-propagating modes, $u_{13}=u_{23}=0$, then
\[
\begin{pmatrix}
z_{11} & z_{12} & z_{13}\\
z_{12} & z_{22} & z_{23}\\
z_{13} & z_{23} & z_{33}
\end{pmatrix}
=
\left(
\begin{array}{ccc}
 \dfrac{u_{11}+2 u_{12}+u_{22}}{4 n+4} & \dfrac{u_{11}-u_{22}}{4 \sqrt{n+1}} &
   0 \\
 \dfrac{u_{11}-u_{22}}{4 \sqrt{n+1}} & \dfrac{u_{11}-2 u_{12}+u_{22}}{4} &
   0 \\
 0 & 0 & u_{33} \\
\end{array}
\right)
\]
and $v_3 = u_{33}$. We easily see that $\tanh b = 0$ because $z_{13} = z_{23} =0$. Then $g_{{\color{black}(1,0,0)}} = g_{{\color{black}(0,1,0)}} = 3/8$ and $g_{{\color{black}(1,1,0})} = 1/(n+1)$, as expected.

For small $u_{13}$ and $u_{23}$ we can expand the tunneling exponent $g$ to the lowest order in $u_{13}$ and $u_{23}$. From equations (\ref{supp:appendix_g_1_0}-\ref{supp:appendix-15}) and (\ref{supp:appendix_tanhb}) we find that
the lowest order is the second order. Applying second order perturbation theory to {\color{black} equation (\ref{supp:appendix-20})}, we observe that $v_{3} = u_{33} + O(u_{13}^2{\color{black}+u_{23}^2})$. Therefore
\[
\tanh^2 b=\frac{\left[\left(z_{22}+u_{33}\right) z_{13}-z_{12} z_{23}\right]^2+\left[\left(z_{11}+u_{33}\right) z_{23}-z_{12} z_{13}\right]^2}{\left[\left(z_{11}+u_{33}\right)\left(z_{22}+u_{33}\right)-z_{12}^2\right]^2}+O(u_{13}^4+{\color{black}u_{23}^4}).
\]
Because $\tanh^2 b = O(u_{13}^2+{\color{black}u_{23}^2})$, we can approximate $\sinh^2 b \approx \tanh^2 b$ and $\cosh^2b \approx 1+ \tanh^2 b$. 

Now, when $n=1$ we can calculate $g_{{\color{black}(1,0,0)}}$ and $g_{{\color{black}(0,1,0)}}$ from equation (\ref{supp:appendix_g_1_0}) and equation (\ref{supp:appendix-14}) with equation (\ref{supp:appendix_31}) and equation (\ref{supp:appendix_32}):
\[
g_{{\color{black}(1,0,0)}} =& 
\frac{3 \cosh
   ^2 b+\left[2 \sqrt{2}\sin (2 a) - \cos (2 a)\right]\sinh ^2 b }{8}\\
   \approx&
   \frac{3}{8}+\frac{1}{8}\left[3(x/z)^2+3(y/z)^2+4\sqrt{2}(x/z)(y/z)-(x/z)^2+(y/z)^2\right]\\
   =&\frac{3}{8}+\frac{1}{4}\left(\frac{x}{z}+\sqrt{2}\frac{y}{z}\right)^2;
\]
\[
g_{{\color{black}(0,1,0)}} =& 
\frac{3 \cosh
   ^2 b-\left[2 \sqrt{2}\sin (2 a) + \cos (2 a)\right]\sinh ^2 b }{8}\\
   \approx&
   \frac{3}{8}+\frac{1}{8}\left[3(x/z)^2+3(y/z)^2-4\sqrt{2}(x/z)(y/z)-(x/z)^2+(y/z)^2\right]\\
   =&\frac{3}{8}+\frac{1}{4}\left(\frac{x}{z}-\sqrt{2}\frac{y}{z}\right)^2.
\]
Substituting the expressions for $x/z$ and $y/z$, we get
\[
g_{{\color{black}(1,0,0)}} = 
\frac{3}{8}+\frac{2\left[\left(3 u_{33}+u_{22}\right) u_{13}-\left(u_{33}+u_{12}\right) u_{23}\right]^2}{\det[U_{2\times 2}+u_{33}K_{2\times 2}]^2}
\]
and
\[
g_{{\color{black}(0,1,0)}} = 
\frac{3}{8}+\frac{2\left[\left(3 u_{33}+u_{11}\right) u_{23}-\left(u_{33}+u_{12}\right) u_{13}\right]^2}{\det[U_{2\times 2}+u_{33}K_{2\times 2}]^2},
\]
where
\[
U_{2\times 2} = 
\begin{pmatrix}
    u_{11}&u_{12}\\
    u_{12}&u_{22}
\end{pmatrix}, \quad
K_{2\times 2} = 
\begin{pmatrix}
    m&n\\
    n&m
\end{pmatrix} = \begin{pmatrix}
    3&1\\
    1&3
\end{pmatrix}.
\]

When $n>1$ we calculate $g_{{\color{black}(1,1,0)}}$, which is simpler to obtain than $g_{(1,0,0)}$ as we notice that
\[
\cos^2 a \sinh^2 b \approx \cos^2 a\tanh^2 b = (x/z)^2
\]
according to equation (\ref{supp:appendix_31}). This can be expanded into 
\[
(x/z)^2=\frac{(n+1)\left[(2u_{33}-u_{12}+u_{22})u_{13}+(2u_{33}-u_{12}+u_{11})u_{23}\right]^2}{\det[U_{2\times 2}+u_{33}K_{2\times 2}]^2}.
\]
Therefore
\[
g_{{\color{black}(1,1,0)}} = \frac{1}{n+1}+\frac{2\left[(2u_{33}-u_{12}+u_{22})u_{13}+(2u_{33}-u_{12}+u_{11})u_{23}\right]^2}{\det[U_{2\times 2}+u_{33}K_{2\times 2}]^2},
\]
where
\[
U_{2\times 2} = 
\begin{pmatrix}
    u_{11}&u_{12}\\
    u_{12}&u_{22}
\end{pmatrix}, \quad
K_{2\times 2} = 
\begin{pmatrix}
    m&n\\
    n&m
\end{pmatrix} = \begin{pmatrix}
    n+2&n\\
    n&n+2
\end{pmatrix}.
\]

{
\color{black}
\section{Equilibration between the fractional and integer channels in the PH-$mmn$ state}

\begin{figure}
    \centering
    \includegraphics[width=0.5\linewidth]{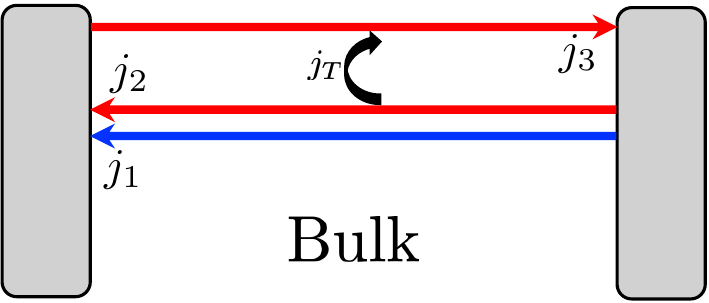}
    \caption{Inter-channel tunneling between the spin-down fractional mode and the spin-down integer mode.}
    \label{supp:fig:interchannel}
\end{figure}

Consider an edge between terminals with potentials $V_{R,L}$ on the right and left (Fig. 3).
Assuming the positive current direction to the left, one expresses the local currents in the three modes in terms of the local voltages $V_i(x)=\mu_i(x)/e$
as

\begin{equation}
\label{supp:A.5.1}
h(j_1, -j_2,-j_3)=e^2(V_1,-V_2, V_3)K^{-1}.
\end{equation}
Spin conservation implies the conservation of the current $j_1$ along the edge. The equilibration of the second and third channels happens due to the tunneling current we take as $j_T(x)=e^2(V_2(x)-V_3(x))/h(m-1)\ell_{\rm eq}$ in the simplest model, where $\ell_{\rm eq}\ll L$ is the equilibration length and $L$ is the edge length. Solving equation (\ref{supp:A.5.1}) together with the continuity equations $\partial_x j_{2,3}=j_T$ and the boundary conditions $V_{1,2}(x=L)=V_R$, $V_3(0)=V_L$ shows the voltage equilibration: 

\begin{equation}
\label{supp:A.5.2}
V_2(x)=V_L+\left(V_R-V_L\right) \frac{m e^{x / \ell_{\mathrm{eq}}}-1}{m e^{L / \ell_{\mathrm{eq}}}-1},
\end{equation}
so that $V_2(0) = V_L$ for $\ell_\mathrm{eq}/L\ll 1$. Interestingly, $V_1(0)=(1+n/m)V_R-nV_L/m$. The currents follow the equations

\[
\left\{
\begin{aligned}
j_1(x) & =\frac{e^2}{h} \frac{V_R}{m-n}, 
\\
j_2(x) & =\frac{e^2}{h}\left[\frac{V_R}{m-n}-\left(V_R-V_L\right) \frac{e^{L / \ell_{\mathrm{eq}}}-e^{x / \ell_{\mathrm{eq}}}}{m e^{L / \ell_{\mathrm{eq}}}-1}\right] ,
\\
j_3(x) & =\frac{e^2}{h}\left[V_L+\left(V_R-V_L\right) \frac{e^{x / \ell_{\mathrm{eq}}}-1}{m e^{L / \ell_{\mathrm{eq}}}-1}\right] .
\end{aligned}
\right.
\]
Setting $V_L=0$, we get $j_2(0)=nj_1(0)/m$. 

This relation also holds for the equilibration process of the two spin-down modes when tunneling currents are injected into the spin-up and -down fractional modes. We thus find that $j_2(0)=nI_\uparrow/m$, and the total drain current on the left is $(n+m)I_\uparrow/m$. The drain current on the right can be found from the charge conservation as $I_{\downarrow}-nI_{\uparrow}/m$. This establishes the relation between the low-frequency drain and tunneling currents and noises.

}


\begin{thebibliography}{74}
\bibitem [{foo()}]{footnote-neutral}%
  \BibitemOpen
  \bibinfo {title} {\color{black}{At} large $n$, operators transferring
  $(k,k,0)$, $k>1$ quasiparticles have a low scaling dimension. {They} do not
  compete with the (1,1,0) particles since their tunneling amplitudes are much
  lower.}\BibitemShut {Stop}%
\end{thebibliography}
\end{document}